\documentclass[twocolumn,showpacs,a4paper]{revtex4}

\usepackage{mathbbol,amssymb,latexsym,amsfonts,amsmath,amsthm}
\usepackage{graphicx}

\newcommand{\I}{\Eins}
\newcommand{\angst}{\mathring{A}}

\DeclareMathOperator{\acos}{acos}

\begin{document}

\title{A helicoidal transfer matrix model for inhomogeneous DNA
  melting}

\date{\today}

\author{Tom Michoel}
\email{tom.michoel@psb.ugent.be}

\author{Yves Van de Peer}
\email{yves.vandepeer@psb.ugent.be}

%\homepage{http://bioinformatics.psb.ugent.be/}

\affiliation{%
Bioinformatics \& Evolutionary Genomics\\
Department of Plant Systems Biology\\
VIB/Ghent University\\
Technologiepark 927, B-9052 Gent, Belgium}

\begin{abstract}
  An inhomogeneous helicoidal nearest-neighbor model with continuous
  degrees of freedom is shown to predict the same DNA melting
  properties as traditional long-range Ising models, for free DNA
  molecules in solution, as well as superhelically stressed DNA with a
  fixed linking number constraint.  Without loss of accuracy, the
  continuous degrees of freedom can be discretized using a minimal
  number of discretization points, yielding an effective transfer
  matrix model of modest dimension $(d=36)$. The resulting algorithms
  to compute DNA melting profiles are both simple and efficient.
\end{abstract}


\maketitle

\section{Introduction}
\label{sec:introduction}

The computation of the thermal stability and statistical physics of
nucleic acids is a classical problem going back to the 1960's. The
standard model to describe the untwisting and separation of both
strands of a free DNA double-helix in solution is the Poland-Scheraga
helix-coil model, where each base pair can be in two possible states,
helix (closed) or coil (open) \cite{poland1970,wartell1985}. Addition
of entropy weights to a basic Ising model, counting the number of
possible configurations of open loops, induces an effective long range
interaction between base pairs which is essential for correctly
obtaining the helix specific opening probabilities.  The most widely
used algorithm for computing the opening probabilities is the
recursion relation method of Poland \cite{poland1974}.  Incorporating
the Fixman-Freire approximation \cite{fixman1977} for the loop entropy
factor reduces the computational complexity from $\mathcal{O}(N^2)$ to
$\mathcal{O}(N)$ in the sequence length $N$.  With the availability of
fully sequenced genomes, the study of DNA melting or denaturation has
become an active field of research again, with recent results relating
the physics of denaturation to the biology of genomes
\cite{yeramian2000,carlon2004}, reparameterizing the original loop
entropy weights \cite{blossey2003}, speeding up the
Poland-Fixman-Freire algorithm for whole genome sequences
\cite{toestesen2003}, and generalizing the model to describe
hybridization with mismatches of unequal length sequences
\cite{garel2004}. The traditional physics approach to compute
statistical mechanical probabilities by transfer matrix multiplication
\cite{kramers1941,onsager1944} has also recently been revisited by
Poland \cite{poland2004}. While this last algorithm offers no
improvement in computational complexity (using matrix sparsity it is
$\mathcal{O}(N^2)$), it is very simple and straightforward to
implement.

\emph{In vivo} DNA strand separation involves interactions with other
molecules which impose superhelical stresses on the DNA molecule. This
is modeled by Benham's statistical mechanical model for stress induced
duplex destabilization (SIDD) \cite{fye1999}, which also is a
helix-coil model with Ising degrees of freedom. It has a long range
base pair interaction arising through superhelical constraints (no
loop entropy factors are added), and opening probabilities are known
to correlate very well with regions important for transcriptional
regulation \cite{benham1993,benham1996}.  An exact solution of the
model is $\mathcal{O}(N^2)$ but an accelerated algorithm using an
energy cut-off reduces this to $\mathcal{O}(N)$, such that SIDD
properties can be computed for whole genome sequences as well
\cite{bi2004,benham2004}.

In parallel with the helix-coil models, a distinct class of
statistical mechanical models for DNA melting has been developed
starting from a physically more realistic description of a base pair
as an entity which has a continuum of intermediate states in between
helix or coil. These models are all based on the Peyrard-Bishop model
\cite{peyrard1989} which consists of a nonlinear particle lattice with
one real degree of freedom per base pair describing the stretching of
the hydrogen bonds between the bases. Nonlinearity and cooperativety
in such a model arises already with a nearest-neighbor interaction, no
long-range interaction is needed \cite{peyrard2004}.  Subsequent
improvements to the model include replacing the harmonic by an
anharmonic stacking energy \cite{dauxois1993}, and introducing an
additional angular degree of freedom per base pair to model the
helicoidal structure of DNA \cite{barbi1999,cocco1999,barbi2003}. In
the latter model, separation of the two strands is coupled to
untwisting of the double helix. The effect of sequence inhomogeneity
on the melting transition in the Peyrard-Bishop model with harmonic
and anharmonic stacking has been studied for random sequences
\cite{cule1997} and for periodic sequences \cite{zhang1997}.

Recent experimental developments (see \cite{strick2003} for a review)
have made it possible to manipulate single polymeric molecules
directly and thus offer access to a whole new range of DNA properties
other than the melting phenomenon. These elasticity experiments too
can be accurately modeled by yet another type of statistical
mechanical models consisting of a double-helix with non-opening base
pairs connected by flexible, folding backbones \cite{zhou1999,
  zhou2000}. In this paper however, we will be concerned with the
melting transition only and the connection between the continuous
particle-lattice models and the discrete helix-coil models.

Unlike the helix-coil models, which have seen many applications to
real biological sequences, the particle-lattice models are mostly used
to obtain a more fundamental, sequence independent, physical
understanding of the DNA melting phenomenon, such as the order of the
phase transition, the existence of nonlinear `bubble' excitations,
etc. (see \cite{peyrard2004} for a recent review paper). Moreover,
although both types of models have been validated against (different)
experiments, very little is known about how they relate to one another
and whether they are in some sense equivalent. Here, we attempt to
close the gap between both kinds of models. We study an inhomogeneous
particle-lattice model based on the Barbi-Cocco-Peyrard helicoidal
model \cite{barbi1999} and compute its melting properties for some
standard example sequences both under free conditions and with
superhelical stresses.

The thermally induced melting of free DNA is obtained as the formally
very simple transfer integral equilibrium solution of the helicoidal
model, yet computation of the melting properties is a challenge in
itself as, e.g., a computation of the partition function involves
$\mathcal{O}(N)$ numerical integrations over an infinite integration
domain.  However, Zhang et al.  \cite{zhang1997} already observed that
for the Dauxois-Peyrard-Bishop model \cite{dauxois1993}, the numerical
integrations can be carried out using a very limited number of
discretization points: a dimension as small as $d=70$ gave very
accurate results compared to much higher dimensions ($d=800$ and
more), and by allowing an error of order $10^{-6}$ with respect to the
exact results, the dimension could be further reduced to $d\approx
40$. For the helicoidal model, we have found a value of $d=36$ to be
the minimal discretization dimension.  This effectively reduces the
particle-lattice model to a nearest-neighbor generalized Ising model,
offering the possibility to develop a very simple and very fast
algorithm to compute DNA melting probabilities. We propose such an
algorithm which moreover is numerically stable for arbitrary sequence
lengths, avoiding underflow problems related to the extensivity of the
free energy (i.e., the exponential vanishing of the partition function
for diverging sequence length). The algorithm is as simple as Poland's
matrix algorithm \cite{poland2004} and as fast as any of the fastest
helix-coil algorithms discussed above.  Extension of the algorithm to
compute correlations between different base pairs, loop opening
probabilities, higher order moments for base pair opening, etc. is
trivial and straightforward.

Stress induced DNA melting is modeled by imposing a fixed linking
number constraint on the DNA strands, which leads to a coupling of all
angular degrees of freedom in the model. However, the linking number
is thermodynamically conjugated to an external torque variable applied
at both ends of the molecule. The model with external torque can be
solved by the above transfer matrix algorithm, and although there is
no equivalence of ensembles, the fixed linking number solution can be
obtained by a complex integration over the torque variable. The
numerical solution is $\mathcal{O}(M\, N)$ where $M$ is a constant
independent of $N$ determined by the desired accuracy of the torque
integration, a situation similar to the analysis of stress induced DNA
melting using Benham's SIDD model \cite{benham2004}.

\section{The model and its equilibrium solution} 
\label{sec:model}

We consider the helicoidal model introduced by Barbi, Cocco and
Peyrard \cite{barbi1999}.  Unlike the original homogeneous model, the
various energy parameters will be explicitly sequence dependent. A DNA
sequence is a string of $N$ letters $\{A,C,G,T\}$, which for
convenience we translate (alphabetically) into a numerical sequence
$(s_n)_n$ taking values in $\{1,2,3,4\}$.  Each base pair in the model
has two degrees of freedom, a radial variable $r$, related to the
opening of the hydrogen bonds, and an angular variable $\phi$, related
to the twisting of the base pair and responsible for the
$3$-dimensional structure of the DNA molecule.  Successive angles are
restricted to $\phi_{n+1} - \phi_n\in [0,\pi]$ to enforce helical
geometry. Alternatively, we can associate a radial variable $r$ to the
sites of the lattice, and an angular variable $\theta\in[0,\pi]$ to
the bonds of the lattice ($\theta_n = \phi_{n+1} - \phi_n$).

The potential energy is given by
\begin{equation}\label{eq:1}
  \begin{split}
    V &= \sum_{n=1}^N D_{s_n} \bigl( e^{-a_{s_n}(r_n-r_0)} -1 \bigr)^2\\
    &\quad+ \sum_{n=1}^{N-1} K_{s_n,s_{n+1}} (r_{n+1}-r_n)^2
    e^{-\alpha(r_n+r_{n+1} -2r_0)}\\
    &\quad+ \sum_{n=1}^{N-1} E_{s_n,s_{n+1}}\bigl(\ell_{n,n+1}-
    \ell^{(0)}_{s_n,s_{n+1}})^2 - \Gamma \sum_{n=1}^{N-1} \theta_n .  
  \end{split}
\end{equation}
The first term is the Morse potential modeling the hydrogen bonds
between the two nucleotides in a base pair \cite{peyrard1989}, the
second term is the anharmonic stacking interaction between successive
base pairs \cite{dauxois1993}, the third term is a harmonic twist
energy allowing fluctuations of the length $\ell_{n,n+1}$ between
successive nucleotides on the same DNA strand \cite{barbi1999} (see
also Appendix \ref{sec:energy-parameters}), and the last term, which
can be written as $-\Gamma(\phi_N-\phi_1)$, is the external torque or
superhelical twist. $\Gamma>0$ overtwists the DNA-molecule, inhibiting
denaturation of the two strands, $\Gamma<0$ causes undertwisting and
enhances denaturation \cite{barbi2003}.

We note here that the present model is only suitable for negative or
small positive torque $\Gamma$. Indeed, unwinding leads to
denaturation and this is well described by the potential energy
(\ref{eq:1}), but severe overwinding leads to new DNA forms with
exposed bases and the backbone winding at the center
\cite{strick2003}. Such transition can obviously not be part of the
present model.

A variety of boundary conditions (b.c.) can be considered for the
radial variable $r$, such as free b.c., fixed b.c., or periodic b.c.,
with minor modifications to the numerical solution of the model.  For
the angular variable $\phi$ we consider two distinct situations. The
first is to set $\phi_1=0$ and have no constraint on $\phi_N$,
corresponding to free b.c. for the variables $\theta$, and describing
the situation in some single molecule experiments \cite{strick2003}.
The second situation, modeling superhelical stresses, is to set a
fixed linking number constraint $\phi_N-\phi_1=\sum_n\theta_n = \alpha
N$, $\alpha\in[0,(N-1)\pi/N]$, which contains periodic b.c. in $\phi$ as
the special case $\alpha=2\pi n/N$, $n=1,2,3\dots$ The torque $\Gamma$
and the total twist $\sum_n\theta_n$ are thermodynamically conjugated
variables, yet as we are explicitly working with a finite-size system,
there is no equivalence of ensembles and both situations lead to
different melting properties. We will refer to the first situation as
the \emph{`torque ensemble'} and the second as the
\emph{`linking number ensemble'}.

The choice of the energetic parameters is a difficult one and unlike
for the helix-coil models, no well established set of parameters
exists, especially with respect to the base pair dependence of the
different energy terms.  Morse potential constants for weakly bonded
$A-T$ vs.  strongly bonded $C-G$ base pairs have been determined by
comparison of the Dauxois-Peyrard-Bishop model with denaturation
experiments \cite{campa1998}. For the other parameters, we follow the
classification of El Hassan and Calladine \cite{elhassan1997}. More
precisely we take $K_{s,t}$ inversely proportional to the slide
variance of the step $(s,t)$, and $E_{s,t}$ inversely proportional to
the twist variance \cite[Table 2]{elhassan1997}. To obtain explicit
values, we first adapt the relative strength of the energy parameters
such that their order of magnitude agrees with the parameters used in
\cite{barbi2003}. In that case we obtain the correct transition
temperature interval, but a less perfect differential melting map (a
melting map gives for each base pair the temperature at which it
transforms from closed to open, see \cite{blake1999} and Section
\ref{sec:res_tq}). By increasing the relative strength of the twist
energy, the transition interval is widened, but the melting map
becomes exact. To compute opening probabilities and melting maps, and
identify stable vs.  unstable regions, this last set of parameters is
more adequate.  A more detailed comparison with experiment will be
needed to find the parameters which best fit the physical melting
transition, but we do not pursue this further in this paper. All the
explicit numerical values we use are given in Appendix
\ref{sec:energy-parameters}. To conclude, we mention that in the
torque ensemble, to first order, sequence specificity in the
melting process comes from the base pair specific Morse potentials,
but inhomogeneity in the stacking and twist energies has second order
effects which are nonetheless important for a detailed identification
of the different melting domains. In the linking number
ensemble, the coupling of all angular degrees of freedom leads
to more complicated sequence specific melting behavior.

\subsection{Equilibrium solution in the torque ensemble}
\label{sec:torque}

Since we are not interested in velocity dependent quantities, the
kinetic energy terms can be integrated directly in the partition
function, which becomes, up to a multiplicative constant and with free
b.c.,
\begin{equation}\label{eq:7}
  Z = \int dr_1\dots\int dr_{N}\int d\theta_1\dots\int d\theta_{N-1}\;
  r_1\dots r_{N} \; e^{-\beta V}.
\end{equation}
The $\theta$-integrals factorize, and
\begin{equation*}
  Z = \int dr_1 \dots\int dr_{N}\, T^{(1)}(r_1,r_2)\dots 
  T^{(N-1)}(r_{N-1},r_N),
\end{equation*}
where for $n=1,\dots, N-2$,
\begin{multline*}
  T^{(n)}(r,r') = r e^{-\beta V^{(n)}_m(r)} e^{-\beta V^{(n)}_s(r,r')}\\
  \times \int_{-1}^1 \frac{dx}{\sqrt{1-x^2}} e^{-\beta V^{(n)}_t(r,r',x)} 
  e^{\beta\Gamma\acos(x)},
\end{multline*}
and
\begin{multline*}
  T^{(N-1)}(r,r') = rr' e^{-\beta[ V^{(N-1)}_m(r) + V^{(N)}_m(r')]}
  e^{-\beta V^{(N-1)}_s(r,r')}\\ \times \int_{-1}^1
  \frac{dx}{\sqrt{1-x^2}} e^{-\beta V^{(N-1)}_t(r,r',x)}
  e^{\beta\Gamma\acos(x)}.
\end{multline*}
$V^{(n)}_m$, $V^{(n)}_s$, and $V^{(n)}_t$ denote respectively the
Morse, stacking, and twist energy terms. As we will not need spectra
of transfer integral operators, there is no need for symmetrizing
these kernels.

In order to compute expectation values of the form $\langle f(r_n)
g(\cos\theta_n)\rangle$ for suitable test functions $f$ and $g$, we
need additional transfer integral operators
\begin{multline*}
  T^{(n)}_{f,g}(r,r') = r f(r) e^{-\beta V^{(n)}_m(r)} e^{-\beta
    V^{(n)}_s(r,r')}\\
  \times\int_{-1}^1 \frac{dx}{\sqrt{1-x^2}} g(x) e^{-\beta
    V^{(n)}_t(r,r',x)}  e^{\beta\Gamma\acos(x)},
\end{multline*}
with appropriate modifications for the right-most sites $N-1$ and $N$.

Since strand separation and untwisting are directly correlated by the
twist energy term, it is often sufficient to consider the case
$g\equiv 1$, such that we get the simpler kernels
\begin{align}
  T^{(n)}_{f}(r,r') &= f(r)\,T^{(n)}(r,r')\label{eq:2}\\
  T^{(N)}_{f}(r,r') &= T^{(N-1)}(r,r')\, f(r').\label{eq:3}
\end{align}

To solve the model numerically, we replace the transfer integral
operators by finite size transfer matrices. The most efficient way for
doing this is approximating the integrals in the partition function by
finite sums using Gaussian quadratures \cite{press1992}.  For the
angular $x$-integrals, this is straightforward as they already contain
the right weight function for Gauss-Chebyshev integration. For the
radial $r$-integrals, we first restrict the infinite integration
domain to a finite interval $[a,b]$, then apply Gauss-Legendre
integration.  Let $z_j$, $j=1,\dots,M_C$, be the zeros of the $M_C$'th
Chebyshev polynomial, all having equal weight $\pi/M_C$. Let $z'_j$,
$j=1,\dots,M_L$, be the zeros of the $M_L$'th Legendre polynomial,
$\xi_j=\frac12(b-a)z'_j+\frac12(b+a)$ the zeros transformed to the
interval $[a,b]$, and $w_j$ the associated weights \cite{press1992}.

We obtain the transfer matrix approximation to the partition function,
\begin{align*}
  Z = \sum_{i,j} \bigl( \hat T^{(1)}\dots \hat T^{(N-1)}\bigr)_{ij}
  = \langle v | \hat T^{(1)}\dots \hat T^{(N-1)} |v\rangle,
\end{align*}
where $v = (1\;1\;\dots\;1)$, $|\cdot\rangle$ and $\langle\cdot|$
are the familiar Dirac column, resp. row vector notation, and $\hat T^{(n)}$
are the $M_L\times M_L$ transfer matrices defined by
\begin{align*}
  \hat T^{(n)}_{ij} &= w_i \xi_i e^{-\beta V^{(n)}_m(\xi_i)} e^{-\beta
    V^{(n)}_s(\xi_i,\xi_j)}\\
  &\qquad\times\frac\pi{M_C}\sum_{k=1}^{M_C} e^{-\beta
    V^{(n)}_t(\xi_i,\xi_j,z_k)} e^{\beta\Gamma\acos(z_k)}\\
  \hat T^{(N-1)}_{ij} &= w_i w_j \xi_i \xi_j e^{-\beta[
    V^{(N-1)}_m(\xi_i) + V^{(N)}_m(\xi_j)]} e^{-\beta
    V^{(n)}_s(\xi_i,\xi_j)}\\
  &\qquad\times\frac\pi{M_C}\sum_{k=1}^{M_C} e^{-\beta
    V^{(n)}_t(\xi_i,\xi_j,z_k)} e^{\beta\Gamma\acos(z_k)}.
\end{align*}
Likewise matrices $\hat T^{(n)}_{f,g}$ are defined as finite
approximations to the corresponding kernels.

Different boundary conditions in the radial variable can be easily
accommodated by changing the vector $v$: for fixed b.c.
$r_1=r_N=\xi_j$, $v_i=\delta_{ij}$, for closed, resp. open b.c., $v_i
= I(\xi_i\leq12)$, resp. $v_i=I(\xi_i>12)$, and for periodic b.c. the
inner product $\langle v|\cdot|v\rangle$ is replaced by the trace
$\mathrm{Tr}(\cdot)$. Here we follow the convention that a base pair
is `open' if $r-r_0>2\angst$, and $I$ denotes the indicator function,
$I(A)=1$ if the condition $A$ is true.

Defining left and right matrix products
\begin{align*}
  M^{(n)}_L  = \hat T^{(1)}\dots \hat T^{(n)}, &&
  M^{(n)}_R  = \hat T^{(n)}\dots \hat T^{(N-1)}
\end{align*}
for $n=1,\dots,N-1$, and $M^{(0)}_L=M^{(N)}_R=\I$, we obtain
\begin{align*}
  \langle f(r_n)g(\cos\theta_n)\rangle &= \frac{\langle v |
    M_L^{(n-1)} \hat T^{(n)}_{f,g} M_R^{(n+1)}| v\rangle}Z\\
  \langle f(r_{N})\rangle &= \frac{\langle
    v | M_L^{(N-2)} \hat T^{(N)}_{f} | v\rangle}Z.
\end{align*}

The different transfer matrices $\hat T^{(n)}$ for $n=1,\dots,N-2$
choose between 16 different matrices, one for each nucleotide step
type. These matrices, together with one matrix $\hat T^{(N-1)}$ for
the final bond, are computed first and stored on disk. For a given
sequence we then compute and store the left and right matrix products.
For a given pair $(f,g)$ we compute again first the $16$ possible
matrices $\hat T^{(n)}_{f,g}$ and the two matrices $\hat
T^{(N-1)}_{f,g}$ and $\hat T^{(N)}_{f,g}$.  With these matrices, we
can then compute, e.g., a profile $n\mapsto\langle
f(r_n)g(\cos\theta_n)\rangle$ by the above formulas. By the simplicity
of the transfer matrix formalism, the computational complexity of this
procedure clearly increases only linearly with $N$.

However, even for sequences of moderate length (a few kbp with double
precision calculations), the left and right matrix products have such
small entries, that they consist of round-off error only, and the
computations become meaningless. This is a common problem and due to
the extensivity of the free energy.  To make this computation work for
sequences of arbitrary length, we define normalized left and right
vectors:
\begin{align*}
  \langle w^{(n)}_L| = \frac{\langle w^{(n-1)}_L | \hat
    T^{(n)}}{\bigl\| \langle w^{(n-1)}_L | \hat T^{(n)}\bigr\|}, &&
  |w^{(n)}_R\rangle = \frac{\hat T^{(n)} |w^{(n+1)}_R\rangle}{\bigl\|
    \hat T^{(n)} |w^{(n+1)}_R\rangle\bigr\|}
\end{align*}
with $w_L^{(0)}=w_R^{(N)}=v/\|v\|$, and while inductively creating
these vectors we store
\begin{equation*}
  c_n = \bigl\| \hat T^{(n)} |w^{(n+1)}_R\rangle \bigr\|.
\end{equation*}
A short calculation reveals that
\begin{align*}
  \langle f(r_n)g(\cos\theta_n)\rangle &= \frac{\langle w^{(n-1)}_L |
    \hat T^{(n)}_{f,g}| w^{(n+1)}_R\rangle}{c_n \langle w^{(n-1)}_L |
    w^{(n)}_R\rangle}\\
  \langle f(r_N)\rangle &= \frac{\langle w^{(N-2)}_L | \hat
    T^{(N)}_{f}| w^{(N)}_R\rangle}{c_{N-1} \langle w^{(N-2)}_L |
    w^{(N-1)}_R\rangle},
\end{align*}
involving only normalized vectors, sequence length independent
$(f,g)$-matrices, and the constants $c_n$, which are formed by
sequence length independent matrices acting on normalized vectors.

If $g\equiv 1$, transfer matrices are of the form
(\ref{eq:2})--(\ref{eq:3}), and the formulas are even simpler. Denote
by $D_f$ the multiplication operator with the function $f$ and by
$\hat D_f$ its diagonal matrix discretization. We get
\begin{align}
  \langle f(r_n)\rangle &= \frac{\langle w^{(n-1)}_L | \hat D_f|
    w^{(n)}_R\rangle}{\langle w^{(n-1)}_L |
    w^{(n)}_R\rangle}\label{eq:4}\\
  \langle f(r_N)\rangle &= \frac{\langle w^{(N-2)}_L | \hat T^{(N-1)}
    \hat D_{f}| w^{(N)}_R\rangle}{\langle w^{(N-2)}_L | w^{(N-1)}_R\rangle}.
  \label{eq:5} 
\end{align}

This method can be easily extended to compute higher moments. For
example, to compute $\langle f(r_n) f(r_m)\rangle$ for fixed $n$ and
all $m$, we define $\hat T^{(n)'} = \hat T_f^{(n)}$ and $\hat T^{(l)'}
= \hat T^{(l)}$ for $l\neq n$. Writing $\langle\cdot\rangle'$ to
denote expectation with respect to these new transfer matrices, we
have, for functions $f>0$,
\begin{equation}\label{eq:6}
  \langle f(r_n) f(r_m)\rangle = \langle f(r_n)\rangle\, \langle
  f(r_m)\rangle'.
\end{equation}

The practical applicability of the method clearly relies on the grid
size values $M_L$ and $M_C$, which were determined as follows.  First
we started from the value $M_L=70$, which according to Zhang et.  al.
\cite{zhang1997} gives exact results for the Peyrard-Bishop model.
For this value, the upper limit of the integration domain has to be
set to $b=40$, larger values of $b$ require larger $M_L$
\cite{zhang1997}. The lower limit $a$ can be put equal to $9.7$ as the
Morse potential can be considered infinite for smaller values. We
determined a value $M_C=35$ to give accurate results in comparison
with the MELTSIM program \cite{blake1999}.  Like in the Peyrard-Bishop
model \cite{zhang1997}, we then found that $M_L$ could be further
decreased with negligible error, to a value of $36$.  Further reducing
the dimension leads to a dramatic change where suddenly all the
interesting transitional behavior is lost, the chain is either
completely open, or completely closed.  After $M_L$ was minimized, we
decreased $M_C$.  Around $M_C=20$, we loose again all interesting
behavior, but the transition is less sharp in this case.  We settled
on $M_C=24$.

The computational method presented so far works well up to a certain
sequence length, where memory becomes the bottleneck instead of CPU
speed (around $10^6$ bp on a typical PC).  To treat even longer
sequences, the sequence is divided into a number of smaller
overlapping subsequences and the probability profiles of those are
combined to obtain the full length profile. This is a standard
procedure \cite{benham2004,poland2004}, which however is much simpler
in a nearest neighbor model than in the long range helix-coil models.
More precisely, assume we cut the sequence of $N$ base pairs into
$N/N_0$ subsequences of length $N_0$. To correct for the artificial
boundaries thus introduced, we compute the opening probabilities for
an interval $[l N_0 - d, (l+1)N_0 +d]$ but only keep the values for
the interval $[lN_0, (l+1)N_0]$. If $d$ is much larger than the
typical \emph{correlation} length, this gives the exact opening
probability for the full sequence. In the helix-coil model, such a cut
is never exact because $d$ is always smaller than the
\emph{interaction} length. In Section \ref{sec:res_tq}, we will see
that at typical values of $T$ and $\Gamma$ (i.e., values
differentiating between stable and unstable regions), the correlation
length is typically rather short, a few 100 base pairs at most.
Therefore, a window size of length $N_0=10^5$ and overlap $2d$ between
$10^3$ and $10^4$ leads to an exact algorithm for long sequences whose
speed is only mildly affected by the windowing procedure.

\subsection{Equilibrium solution in the linking number ensemble}
\label{sec:lk_num}

The partition function in the linking number ensemble is again
given by an integral of the form (\ref{eq:7}), but the angular
integrals are now restricted to the subspace of $[0,\pi]^{N-1}$ for
which the linking number or total twist satisfies
\begin{equation*}
  \frac1N\sum_{n=1}^{N-1}\theta_n = \alpha
\end{equation*}
for some fixed $\alpha\in[0,(N-1)\pi/N]$. Very often, instead of
$\alpha$, the superhelical density $\sigma$ is specified,
\begin{align*}
  \sigma = \frac{Lk - Lk_0}{Lk_0}
\end{align*}
where $Lk = (2\pi)^{-1}\sum_n\theta_n$ is the linking number, and
$Lk_0 = (2\pi)^{-1}\sum_n \theta^{(0)}_{n,n+1}$ is the ground state,
zero torque linking number. As we remarked in Section \ref{sec:model},
our model is only suitable for negative or small positive
superhelicity, and for definiteness we will consider in this section
only negative torque $\Gamma$, corresponding to negative superhelicity
$\sigma$.

The situation is completely analogous to the standard statistical
mechanics situation of canonical and grand-canonical ensembles:
$\alpha$ plays the role of the `density', $\Gamma$ the role of a
`chemical potential', and by changing in eq. (\ref{eq:7}) the angular
integration variables to
$\theta_1,\dots,\theta_{N-2},\lambda=\sum_n\theta_n$, we see that the
grand-canonical (torque ensemble) and canonical (linking number
ensemble) partition functions are related by a Laplace transform
\begin{equation*}
  Z_{tq}(\Gamma) = \int_0^{\infty} d\lambda\; e^{\beta \Gamma \lambda}
  Z_{lk}(\frac\lambda N),
\end{equation*}
where it is understood that $Z_{lk}(\alpha)=0$ for
$\alpha>(N-1)\pi/N$.

By standard inverse Laplace transform techniques, the linking number
partition function can be obtained from the torque partition function
by a contour integration in the complex plane
\begin{align}
  Z_{lk}(\alpha) &= \beta \int_{\Gamma-i\infty}^{\Gamma+i\infty} 
  \frac{dz}{2\pi i}\; e^{-\beta z N\alpha} Z_{tq}(z) \nonumber\\
  & = \beta \int_{\Gamma-i\infty}^{\Gamma+i\infty} 
  \frac{dz}{2\pi i}\; e^{-\beta N[ z\alpha + F_{tq}(z)]}\label{eq:12}
\end{align}
where the integral is carried out on a line parallel to the imaginary
axis, with $\Gamma<0$, and $F_{tq}(z)$ is the free energy in the
torque ensemble.

Standard statistical mechanics proceeds by choosing a line which
crosses the real axis at a critical point of the harmonic function
$z\alpha + F_{tq}(z)$. This point is a saddle point and the contour is
a path of steepest descent, such that for large $\beta N$, the
integrand in eq. (\ref{eq:12}) is significantly different from zero in
a small interval near the real axis only. Since we are interested in
inhomogeneous, finite sequences, we do not consider the question of
equivalence of ensembles in the thermodynamic limit.

More precisely, for a given $\sigma<0$ and corresponding $\alpha$, the
function
\begin{equation*}
  \Gamma<0 \mapsto \alpha \Gamma + F_{tq}(\Gamma)
\end{equation*}
attains a maximum at some value $\Gamma_0<0$, namely the value
$\Gamma_0$ for which $\langle \sum_n \theta_n\rangle_{tq,\Gamma_0} =
N\alpha$, where $\langle\cdot\rangle_{tq,\Gamma_0}$ denotes
expectation in the torque ensemble.  The line passing through
$\Gamma_0$ is chosen as the integration contour, and we can write
\begin{multline}\label{eq:9}
  Z_{lk}(\alpha) = \beta e^{-\beta N \Gamma_0 \alpha} e^{-\beta N F_{tq}(\Gamma_0)} \\
  \times \int_{-\infty}^\infty \frac{d\omega}{2\pi} e^{-i\beta\omega
    N\alpha} e^{-\beta N(F_{tq}(\Gamma_0+i\omega)-F_{tq}(\Gamma_0))}.
\end{multline}
Because of the large parameter $\beta N$, the function
\begin{align*}
  \omega \mapsto \bigl|e^{-\beta
    N(F_{tq}(\Gamma_0+i\omega)-F_{tq}(\Gamma_0))}\bigr| 
\end{align*}
is tightly concentrated around $\omega=0$, and the integral can be
restricted to a small interval $[-\epsilon_N,\epsilon_N]$. It is
important to remark that to apply the standard stationary phase
expansion, $\epsilon_N$ would have to be much smaller than $(\beta
N)^{-1/2}$, a condition which is typically \emph{not} fulfilled here.
An efficient method to numerically compute the remaining integral
consists of computing the integrand at a number of points and find a
cubic splines interpolation which can be readily integrated.

The algorithm to compute expectation values proceeds as follows. Like
in the previous section, let $(f,g)$ be single-site test functions and
denote by $Z^{(n)}$ the partition functions obtained by substituting
at position $n$ the transfer matrix $\hat T^{(n)}_{f,g}$ for $\hat
T^{(n)}$. Further denote by $p^{(n)}_{tq}(\Gamma)= \langle
f(r_n)g(\cos\theta_n)\rangle_{tq,\Gamma}$ the expectation value at
torque $\Gamma$ and analogously $p^{(n)}_{lk}(\alpha)$. Recalling that
$p^{(n)}_{tq}(\Gamma)= \exp(-\beta N[F^{(n)}_{tq}(\Gamma) -
F_{tq}(\Gamma)])$, we find
\begin{multline}\label{eq:8}
  p^{(n)}_{lk}(\alpha) = p^{(n)}_{tq}(\Gamma_0)\\ \times \frac{\int
    d\omega\, p^{(n)}_{tq}(\Gamma_0+i\omega) e^{-i\beta N
      \alpha\omega} e^{-\beta N(F_{tq}(\Gamma_0+i\omega) -
      F_{tq}(\Gamma_0))}} {\int d\omega\, e^{-i\beta N \alpha\omega}
    e^{-\beta N(F_{tq}(\Gamma_0+i\omega)) - F_{tq}(\Gamma_0))}}.
\end{multline}
Since the l.h.s. of this equation is obviously real, we take the real
parts of the integrands before numerically computing the integral. The
torque expectation values $p^{(n)}_{tq}(\Gamma)$ are evaluated using
the efficient algorithm of Section \ref{sec:torque}, which can be
easily extended to also return the free energy:
\begin{align*}
  -\beta N F_{tq}(\Gamma) &= \ln \langle v| \hat T^{(1)}\dots \hat
  T^{(N-1)} |v\rangle\\
  &= \ln \langle v| \hat T^{(1)}|v\rangle + \sum_{n=2}^{N-1} \ln
  \frac{\langle v| \hat T^{(1)}\dots \hat T^{(n)} |v\rangle}{\langle
    v| \hat T^{(1)}\dots \hat T^{(n-1)} |v\rangle}\\
  &= \sum_{n=1}^{N-1} \ln \frac{\langle w_L^{(n-1)}| \hat T^{(n)}
    |w_R^{(N)}\rangle}{\langle w_L^{(n-1)}|w_R^{(N)}\rangle}
\end{align*}
Hence the algorithm to compute expectation values for all $n$ in
(\ref{eq:8}) is still $\mathcal{O}(N)$, but $M$ times slower than in
the torque ensemble, where $M$ only depends on the number of
discretization points chosen to compute the $\omega$-integrals. 

To prove equivalence of ensembles between the torque and linking
number ensemble (for the given test functions) in the
thermodynamic limit for homogeneous sequences, we would have to show
that the fraction of the integrals in equation (\ref{eq:8}) tends to
$1$. For finite, inhomogeneous sequences, the non-triviality of this
fraction causes a nonlinear coupling between base pairs that will be
illustrated in Section \ref{sec:res_lk}.

\section{Example results}
\label{sec:results}

\subsection{Torque ensemble}
\label{sec:res_tq}

For easy comparison with the Poland-Scheraga helix-coil model, we show
example results for the PN/MCS13 sequence ($N=4608$) which is the main
example of \cite{blake1999}. This sequence is the pBR322 sequence
\footnote{GenBank/EMBL accession number J01749.\\
  \texttt{http://www.ebi.ac.uk/embl/}} with an insert
$[AAGTTGAACAAAAR]_{17}AAGTTGA$ at position $972$ \cite{blake1998}
($[\dots]_x$ means $[\dots]$ $x$ times repeated). The conclusions
drawn here are equally valid for all other sequences we tested.

In the Peyrard-Bishop and related models a base pair is said to be
denatured when it is stretched more than $2\,\angst$ away from its
equilibrium length of $10\,\angst$, hence the probability of
denaturation is given by
\begin{equation}
  p_n = \langle h(r_n-12)\rangle,
\end{equation}
where $h(r)$ is the Heaviside function, $h=1$ for $r\geq0$ and $0$
otherwise. Notice that we only need the simpler formulas
(\ref{eq:4})--(\ref{eq:5}) to compute melting profiles $n\mapsto
p_n$.

\begin{figure}[ht!]
  \includegraphics[width=\linewidth]{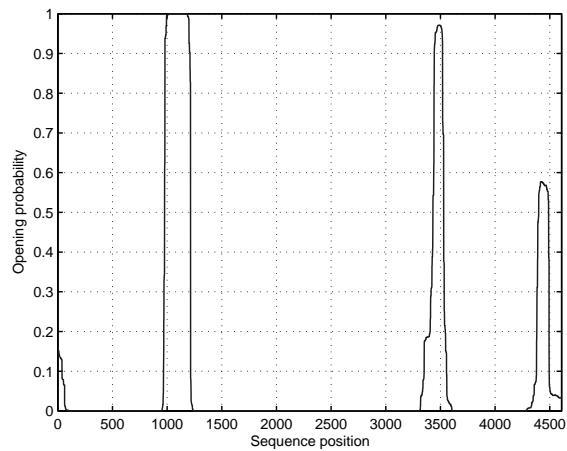}
  \caption{PN/MCS13 Opening probability at $T=310\,K$ and
    $\Gamma = -0.042\,eV/rad$.}
  \label{fig:1}
\end{figure}

Figure \ref{fig:1} shows the melting profile for the PN/MCS13 sequence
at typical \emph{in vivo} temperature $T=310\,K$. The torque value
$\Gamma=-0.042\,eV/rad$ is chosen to give a good delineation of
unstable regions ($p_n\approx 1$). Decreasing $\Gamma$ increases the
number of open base pairs, and increasing $\Gamma$ has the opposite
effect.  On the basis of this melting profile we identify three
unstable regions, the first one around position $1000$ corresponding
to the $AT$-rich insert in the pBR322 sequence, and the other two with
maximums at position $3489$ and at position $4423$.

For whole genome sequences of several million base pairs, the opening
probability is computed by dividing the sequence in shorter
overlapping subsequences (see the end of Section \ref{sec:torque}). To
do this correctly, we need to know the correlation length, or more
precisely the length at which the correlation between a base pair and
the rest of the sequence vanishes. Hence for a fixed base pair $n$ we
compute, using (\ref{eq:6}),
\begin{equation*}
  C^n_m = \langle r_n r_m \rangle - \langle r_n\rangle \langle r_m\rangle.
\end{equation*}

\begin{figure}[ht!]
  \includegraphics[width=\linewidth]{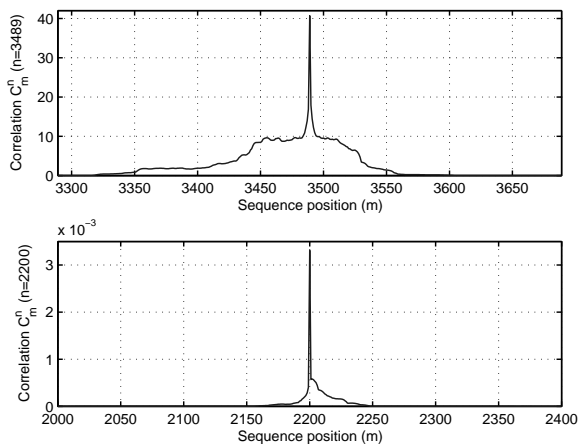}
  \caption{Correlation function $C^n_m$ for $n=3489$ (top) and
    $n=2200$ (bottom) for the PN/MCS13 sequence at $T=310\,K$ and
    $\Gamma = -0.042\,eV/rad$.}
  \label{fig:3}
\end{figure}

Figure \ref{fig:3} shows the correlation function $C^n_m$ for two
different values of $n$, $n=3289$ in the middle of the second opened
bubble (see Figure \ref{fig:1}), and $n=2200$ in the largest closed
region. Clearly, the correlation is much larger in the denatured
region, but even here it does not extend beyond a few 100 base pairs.

Due to the inhomogeneity of base pair bonding and stacking energies,
DNA melting is a stepwise process with different domains melting at
different temperatures. This can be visualized by computing
differential melting curves and melting maps. Let $\gamma$ be the
fraction of open base pairs, $\gamma=(\sum_n p_n)/N$. A differential
melting curve is a plot of $d\gamma/dT$ vs. temperature $T$. A melting
map is obtained by displaying for each temperature the base pairs
which have opening probability greater than $\frac12$ (shaded area).
Such a map gives another picture of thermodynamically stable (high
melting temperature) vs. unstable (low melting temperature) regions in
the particular sequence.

\begin{figure}[ht!]
  \includegraphics[width=\linewidth]{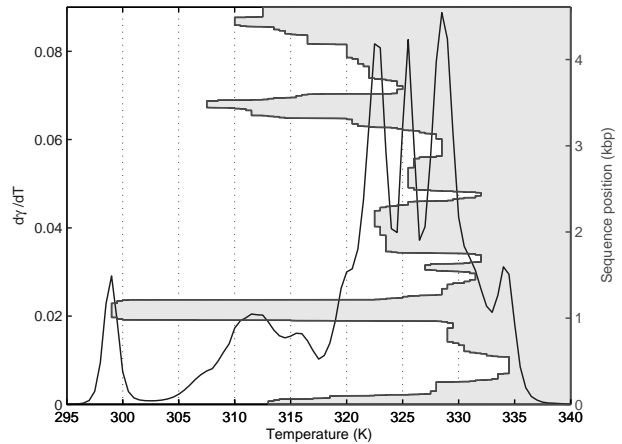}
  \caption{PN/MCS13 Differential melting curve and melting map (shaded
    area) (temperature increment $0.5\,K$, $\Gamma =
    -0.042\,eV/rad$).}
  \label{fig:2}
\end{figure}

Figure \ref{fig:2} shows the differential melting curve (obtained by
differentiating a cubic splines interpolation of the computed values
$\gamma(T)$) and melting map for the PN/MCS13 sequence under the same
condition $\Gamma=-0.042$ as in Figure \ref{fig:1} and \ref{fig:3}.
We can clearly identify again the $AT$-rich inserted region around
position 1000 which melts first, as well as the two unstable regions
around positions 3500 and 4500.

Comparison of the differential melting curve and melting map with the
MELTSIM result \cite[Figure 3]{blake1999} shows first of all that the
general shape of the melting curve is correct, but the temperature
range from a completely closed to a completely denatured molecule is
about twice as large in the helicoidal Peyrard-Bishop model with the
current set of parameters. On the other hand, the melting map as a map
depicting the successive melting order of different regions is in
precise agreement with the MELTSIM melting map.

A more systematic determination of the physical value of the various
energy parameters in the helicoidal model is desirable, but since the
whole process of fitting model computations to experimental results of
DNA denaturation in solution is quite subtle (the experimental results
also depend on external conditions like, e.g., the solvent salt
concentration \cite{poland1970,campa1998}), it falls beyond the scope
of this paper.  It should also be pointed out that while such fitting
was important in the early stages of theoretical study of DNA melting,
present day problems concern more the identification of stable and
unstable regions and linking those to genomic content. As long as the
relative strength of the various energy terms is kept within certain
limits, this identification is unaffected by changing the model
parameters.

\begin{figure}[ht!]
  \includegraphics[width=\linewidth]{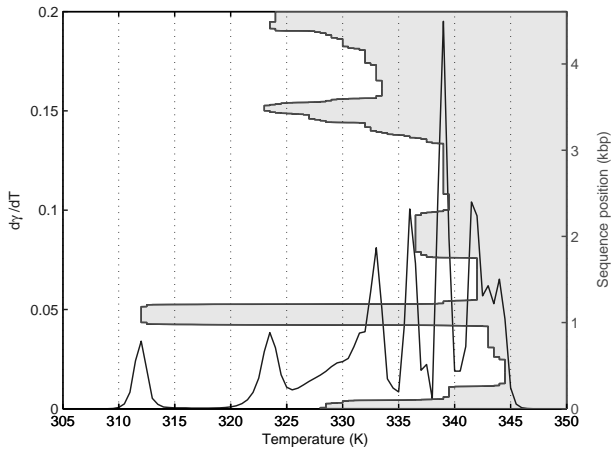}
  \caption{PN/MCS13 Differential melting curve and melting map (shaded
    area) with homogeneous stacking and twist energy ($K =
    0.1486\,eV$, $E = 0.0942\,eV$, $\theta_0 = 34.81^\circ$)
    (temperature increment $0.5\,K$, $\Gamma = -0.042 \,eV/rad$).}
  \label{fig:5}
\end{figure}

\begin{figure}[ht!]
  \includegraphics[width=\linewidth]{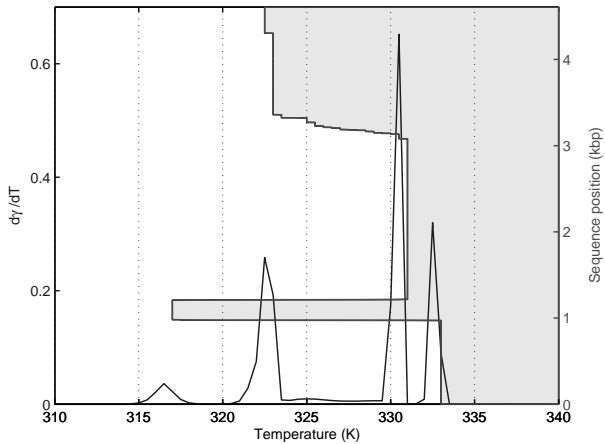}
  \caption{PN/MCS13 Differential melting curve and melting map (shaded
    area) with homogeneous stacking and twist energy ($K = 0.65\,eV$,
    $E = 0.04\,eV$, $\theta_0 = 34.78^\circ$) (temperature increment
    $0.5\,K$, $\Gamma = -0.042 \,eV/rad$).}
  \label{fig:6}
\end{figure}

In Figure \ref{fig:5} and \ref{fig:6} we illustrate some of the
effects of changing the model parameters.  Figure \ref{fig:5} shows
the differential melting curve and melting map with homogeneous
stacking and twist energy terms, where the values of $K$, $E$ and
$\theta_0$ are the averages of the values given in Appendix
\ref{sec:energy-parameters}. The overall identification of stable vs.
unstable regions remains intact, but comparison with Figure
\ref{fig:2} shows that considerable detail in the melting map is lost,
with larger regions melting at once.

Figure \ref{fig:6} shows the effect of changing the relative strength
of the stacking and twisting energy terms. Again they are taken
homogeneous, but now with the original values of Barbi et al.
\cite{barbi2003}. Although with these values the transition
temperature interval is of the right magnitude, the differential
melting curve and melting map clearly display insufficient detail.
Most notably, the two distinct unstable regions around positions
$3500$ and $4500$ are merged into one large region.

\begin{figure}[ht!]
  \includegraphics[width=\linewidth]{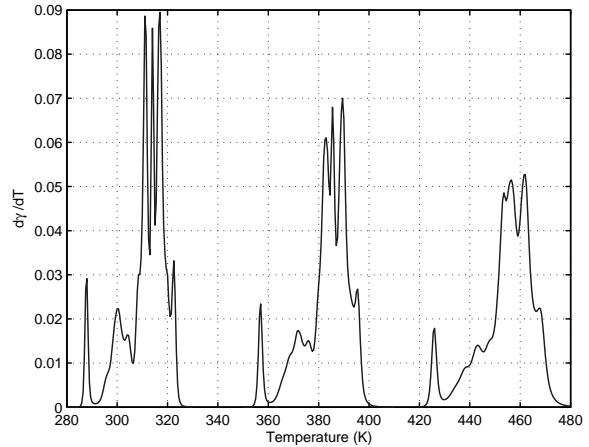}
  \caption{PN/MCS13 Differential melting curves for $\Gamma = -0.05$,
    $0.0$, and $0.05 \,eV/rad$ (left to right, temperature increment
    $0.5\,K$).}
  \label{fig:4}
\end{figure}

So far, we have shown results for a chosen value $\Gamma=-0.042$ for
easy comparison between different figures, but other values can be
considered as well.  At fixed temperature, increasing $\Gamma$
decreases the fraction of open base pairs, corresponding to an
increase in the phase transition temperature in the thermodynamic
limit \cite{barbi2003}.  What is perhaps more interesting is the fact
that increasing $\Gamma$ also smooths the differential melting curve
(see Figure \ref{fig:4}), and broadens the transition; decreasing
$\Gamma$ has the opposite effect.  Heuristically, increasing $\Gamma$
effectively increases the stiffness of the double stranded DNA, which
is indeed known to broaden the transition \cite{carlon2002,
  blossey2003}. The value $\Gamma=0$ plays no special role in this
respect.  In contrast, a recent model \cite{garel2004b} which adds
angular degrees of freedom to the Poland-Scheraga helix-coil model
singles out the value $\Gamma=0$ as special and predicts a broadening
of the transition for $\Gamma<0$ as well as $\Gamma>0$.

Finally, we have also tested the performance of the transfer matrix
algorithm on larger sequences, up to about $N\approx 3\times 10^{5}$.
The algorithm was written in Matlab and run on a 2.8 GHz PC, and
computed times follow the line $t=10^{-4}\,N+0.40$, with $t$ in
seconds.  Comparison with \cite{toestesen2003} shows that our
algorithm performs as fast as the fastest available helix-coil
algorithm.

\subsection{Linking number ensemble}
\label{sec:res_lk}

\begin{figure}[ht!]
  \includegraphics[width=\linewidth]{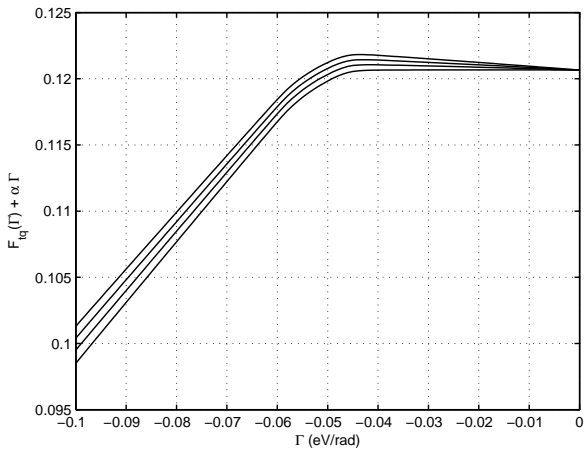}
  \caption{C-MYC $F_{tq}(\Gamma)+\alpha\Gamma$ for $\alpha=0.572$,
    $0.581$, $0.590$, $0.600$ (top to bottom) at $T=310\,K$.}
  \label{fig:7}
\end{figure}

In this section, we illustrate the solution of the fixed linking
number ensemble, and compare it to the fixed torque ensemble as well
as the helix-coil SIDD model \cite{fye1999}, by showing example
results for the C-MYC sequence ($N=3200$), available as Example 3 on
the WebSIDD server \cite{bi2004}. Again, the qualitative conclusions
drawn from this example are valid in general.  Following the outline
of Section \ref{sec:lk_num}, we start by showing in Figure \ref{fig:7}
a plot of $F_{tq}(\Gamma)+\alpha\Gamma$ for different values of the
superhelical density $\sigma=-0.06$, $-0.045$, $-0.03$, $-0.015$,
corresponding to values $\alpha = (1+\sigma)Lk_0/N=0.572$, $0.581$,
$0.590$, $0.600$. As $\sigma$ goes to $0$, the graph becomes constant
for $\Gamma$ smaller than a critical value corresponding to the torque
induced melting transition observed in the homogeneous model
\cite{cocco1999,barbi2003}. For non-zero $\sigma$ the graph has a
maximum at some value $\Gamma_0(\sigma)$ and this is the value we need
for constructing the integration contour and for comparing the linking
number and torque ensembles. We emphasize here that obtaining a very
precise value of the location of the maximum is not necessary. Indeed,
the integration in eq. (\ref{eq:9}) and (\ref{eq:8}) can be carried
out along any line parallel to the imaginary axis. Taking a line at or
close to the maximum will simply ensure that the function to be
integrated falls off very quickly along this line. 

\begin{figure}[ht!]
  \includegraphics[width=\linewidth]{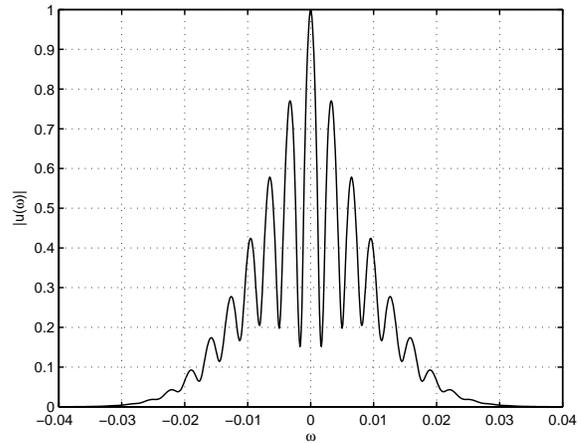}
  \caption{C-MYC Absolute value of $u(\omega)$ (eq. (\ref{eq:10})) at
    $\sigma=-0.03$ ($\alpha=0.59$) and $T=310\,K$ ($\omega$-interval
    $5\cdot 10^{-4}$).}
  \label{fig:8}
\end{figure}

\begin{figure}[ht!]
  \includegraphics[width=\linewidth]{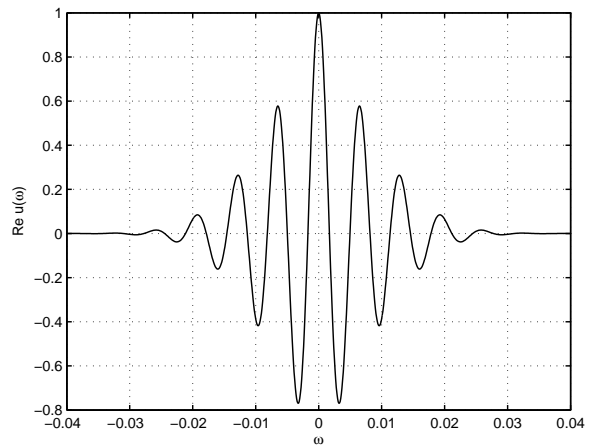}
  \caption{C-MYC Real value of $u(\omega)$ (eq. (\ref{eq:10})) at
    $\sigma=-0.03$ ($\alpha=0.59$) and $T=310\,K$ ($\omega$-interval
    $5\cdot 10^{-4}$).}
  \label{fig:9}
\end{figure}

Next we turn our attention to the integrand in eq.  (\ref{eq:9}),
\begin{equation}\label{eq:10}
  u(\omega) = e^{-\beta N ( F_{tq}(\Gamma_0+i\omega) -
    F_{tq}(\Gamma_0))} e^{-i\beta N\alpha\omega}
\end{equation}
Figure \ref{fig:8} shows the absolute value of $u$ in a neighborhood
of $\omega=0$ for a superhelical density $\sigma=-0.03$ and
temperature $T=310\,K$.  For this value of $\sigma$ and $T$, the
critical point is given by $\Gamma_0=-0.04149$. As expected, the
function decays to $0$ rapidly, but clearly not rapidly enough to
apply a stationary phase approximation ($(\beta N)^{-1/2}=0.003$).
Figure \ref{fig:9} shows the real part of $u$, which is the function
to be integrated to obtain the partition function in eq.
(\ref{eq:9}).  Both Figure \ref{fig:8} and \ref{fig:9} are generated
by interpolating between a number of computed data points. Due to the
oscillations, it is important to compute enough data points, we used
an interval of $\Delta\omega = 5\cdot 10^{-4}$. One way to determine
the accuracy of the numerical approximation is to check if the
expectation value of the superhelical density matches the imposed
value. We find $( (2\pi)^{-1}\sum_n \langle
\theta_n\rangle_{lk,\sigma} - Lk_0)/Lk_0 = -0.03005$ which compares
well with the exact value of $-0.03$. Similarly, we can check how well
the value $\Gamma_0$ was determined by computing the expected helical
density in the fixed torque ensemble with torque $\Gamma_0$, and find
a value of $-0.03026$. This is the value obtained by differentiating
the splines approximation of Figure \ref{fig:7}, and due to the broad
maximum, better accuracy should not be expected. However, it is clear
from the fast decay of the function $|u(\omega)|$ in Figure
\ref{fig:8}, that this value of $\Gamma_0$ is accurate enough for
computing expectation values in the linking number ensemble. If a more
accurate value is desired, such as in Figure \ref{fig:11} below, we
can start from this approximate $\Gamma_0$, and fine tune it by
computing $\langle \sum_n\theta_n\rangle_{tq,\Gamma}$ for some values
$\Gamma\approx\Gamma_0$ until the correct linking number is found.

\begin{figure}[ht!]
  \includegraphics[width=\linewidth]{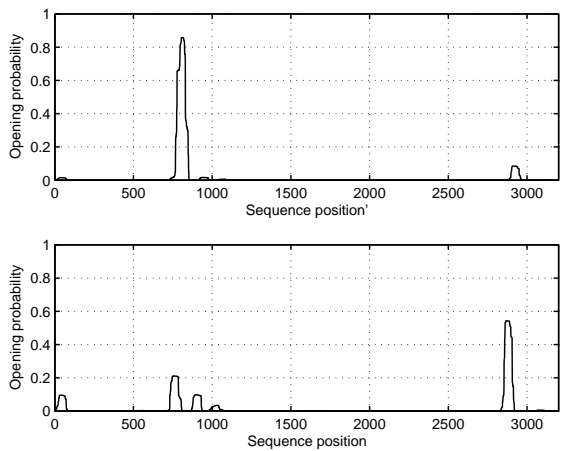}
  \caption{Opening probability at $T=310\,K$ and fixed superhelical
    density $\sigma=-0.03$ for the C-MYC sequence ($N=3200$) (top) and
    the modified C-MYC sequence ($N=3156$) (bottom).}
  \label{fig:10}
\end{figure}

To visualize how the linking number constraint affects the melting
behavior of a sequence by introducing an effective long range base
pair coupling in the partition function, we follow Benham and Bi
\cite{benham2004}, and compare the original C-MYC sequence to a
modified sequence which differs from the C-MYC sequence in a tiny
fragment only. More precisely, we compute the opening probability for
the C-MYC sequence at fixed linking number ($\sigma=-0.03$) (Figure
\ref{fig:10}, top panel), then remove from the sequence a small $44$
bp segment in the center of the main untwisted region (sequence
positions 781 -- 824), and compute the opening probability for this
modified sequence at the same superhelical density (Figure
\ref{fig:10}, bottom panel).

For the C-MYC sequence, we find that there are two locations that are
preferentially opened, a first, large one, between positions $760 -
850$, and a second, smaller one, between position $2900-2950$, with
much higher opening probability for the largest region. In agreement
with the SIDD-model \cite{benham2004} we see that a small modification
of the sequence is sufficient to shift the main opening activity to
the second region.

In Figure \ref{fig:11}, we show the corresponding opening
probabilities in the fixed torque ensemble. For a fair comparison, we
adjusted the torque values for both sequences separately to return a
superhelical density expectation value $\sigma=-0.03$ with high
accuracy.  As a final check that we are comparing both ensembles with
the right parameters, we compute the total fraction of open base pairs
$N^{-1}\sum_n p_n$, and find that it is equal to $0.018$ for both top
panels of Figure \ref{fig:10} and \ref{fig:11}, and equal to $0.017$
for both bottom panels.

The main qualitative difference that can be observed between both
ensembles is that the effect of removing the small segment is much
more localized in the torque ensemble, and the first region is still
dominant. If we consider a base pair to be open if $p_n>0.5$, like in
constructing the melting maps in Section \ref{sec:res_tq}, we see that
in the linking number ensemble the open region shifts from the left to
the right upon modifying the sequence, while in the torque ensemble,
the open region disappears.

\begin{figure}[ht!]
  \includegraphics[width=\linewidth]{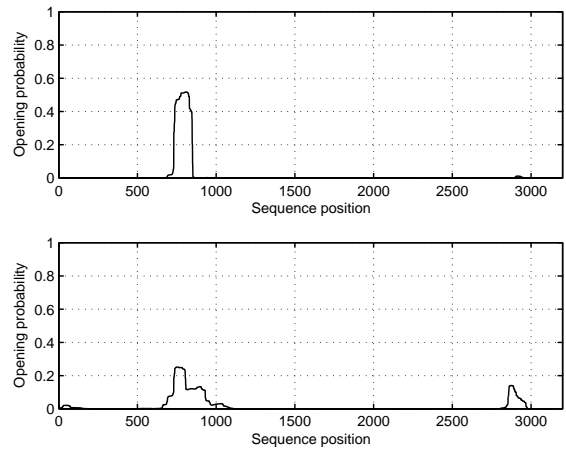}
  \caption{Opening probability at $T=310\,K$ in fixed torque ensemble
    for the C-MYC sequence ($N=3200$, $\Gamma=-0.041475$,
    $\langle\sigma\rangle_{\Gamma} = -0.03004$) (top) and the
    modified C-MYC sequence ($N=3156$, $\Gamma=-0.044365$,
    $\langle\sigma\rangle_{\Gamma} = -0.03003$) (bottom).}
  \label{fig:11}
\end{figure}

\section{Conclusions and outlook}
\label{sec:conclusion-outlook}

In this paper we have connected the particle-lattice helicoidal
Peyrard-Bishop model to more familiar Ising-type models for
inhomogeneous DNA melting. In the simplest setting of a fixed external
torque, the model has the same melting behavior as the Poland-Scheraga
helix-coil model. Since the numerical integrations needed to compute
the profiles in the particle-lattice model can be carried out using a
limited number of discretization points, and since the interactions
are only nearest-neighbor, we have obtained a new method to compute
melting profiles which is both very simple to implement and very
efficient to execute, and which is therefore highly attractive to
analyze very long, or even whole genome sequences.

Furthermore we have shown that also the more complicated setting of a
fixed linking number can be treated and that the results are in
agreement with Benham's SIDD model. The algorithm is again simple to
implement and consists of numerically integrating the fixed torque
results over a small range of complex torque values.  Some of the
points raised here such as the inequivalence of ensembles are
worthwhile of investigating mathematically more rigorous in the setting
of the homogeneous helicoidal Peyrard-Bishop model to see if they
persist in the thermodynamic limit.

The equivalence between nearest-neighbor lattice models with
continuous degrees of freedom on the one hand, and Ising models with
loop entropy weights or long-range interaction on the other hand,
raises more fundamental questions as well. A better understanding of
this equivalence will presumably lead to a better understanding of
nonlinear phenomena in one-dimensional systems.

\begin{acknowledgements}
  We thank an anonymous referee for helpful remarks concerning Section
  \ref{sec:lk_num} and \ref{sec:res_lk}, and a second anonymous
  referee for pointing out reference \cite{zhou1999}.  

  T.M. is a Postdoctoral Fellow of the Research Foundation Flanders
  (F.W.O.--Vlaanderen).
\end{acknowledgements}

\appendix

\section{Energy parameters}
\label{sec:energy-parameters}

In this appendix we collect the various parameters used in the
potential energy (\ref{eq:1}).  All lengths are measured in $\angst$,
energies in $eV$, and angles in $rad$.

For the depth $D_i$ of the Morse potentials, we choose values close to
the value $0.15$ of \cite{barbi2003}, but taking into account that a
$C-G$ base pair has a $1.5$ times stronger bond than an $A-T$ base
pair. For the widths $a_i$ we take the values of \cite{campa1998}.
\begin{align*}
  D_1 &= D_4 = 0.12 & D_2 &= D_3 = 0.18\\
  a_1 &= a_4 = 4.2 & a_2 &= a_3 = 6.9.
\end{align*}
The equilibrium distance $r_0$ is equal to $10$.

The length $\ell_{n,n+1}$ between successive nucleotides on the same
DNA strand in the twist energy term is given by
\begin{equation*}
  \ell_{n,n+1} = \sqrt{h^2+r_n^2+r_{n+1}^2 - 2r_n r_{n+1} 
    \cos\theta_n},
\end{equation*}
where $h=3.4$ is the fixed vertical distance between base
pairs. The rest length $\ell^{(0)}_{n,n+1}$ is step dependent and
given by
\begin{align*}
  \ell^{(0)}_{s_n,s_{n+1}} = \sqrt{h^2 + 4 r_0^2
    \sin^2(\tfrac12\theta^{(0)}_{s_n,s_{n+1}})}
\end{align*}
where $\theta^{(0)}_{s_n,s_{n+1}}$ is the average helical twist angle of the
given step, taken from the database of El Hassan and Calladine
\cite{elhassan1997} 
\begin{align*}
  \theta^{(0)} = \frac{2\pi}{360}\times
  \begin{pmatrix}
    35.9& 32.9& 34.8& 32.4\\
    37.4& 31.9& 35.1& 34.8\\
    37.8& 37.4& 31.9& 32.9\\
    30.6& 37.8& 37.4& 35.9
  \end{pmatrix}.
\end{align*}
The parameter $E$ is taken inversely proportional to the twist angle
standard deviations, taken from the same database \cite{elhassan1997}:
\begin{align*}
  E=0.4\times
  \begin{pmatrix}
    0.3030&    0.2632&    0.2083&    0.3571\\
    0.1053&    0.2703&    0.1887&    0.2083\\
    0.2632&    0.2500&    0.2703&    0.2632\\
    0.1493&    0.2632&    0.1053&    0.3030\\
  \end{pmatrix}.
\end{align*}
Similarly, the stacking energy parameter $K$ is taken inversely
proportional to the slide standard deviations of \cite{elhassan1997}:
\begin{align*}
  K &= 0.1\times
  \begin{pmatrix}
    3.5714&    1.4085&    1.2195&    2.0833\\
    0.8130&    0.8547&    0.9804&    1.2195\\
    1.4493&    1.1628&    0.8547&    1.4085\\
    0.9174&    1.4493&    0.8130&    3.5714
  \end{pmatrix}.
\end{align*}
The constant $\alpha$ in the exponential is put equal to $0.5$ as in
\cite{barbi2003}.

%\bibliography{biblioDNA}

\end{document}